# Recommending Dream Jobs in a Biased Real World

## Nadia FAWAZ
LinkedIn Corporation
nfawaz@linkedin.com

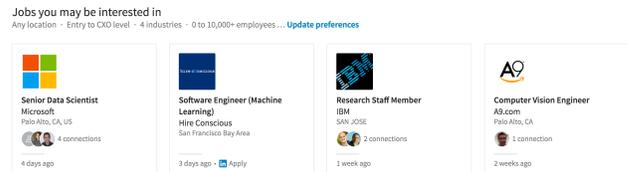

**Figure 1: LinkedIn JYMBII (Jobs You May Be Interested In)**

## ABSTRACT

Machine learning models learn what we teach them to learn. Machine learning is at the heart of recommender systems. If a machine learning model is trained on biased data, the resulting recommender system may reflect the biases in its recommendations. Biases arise at different stages in a recommender system, from existing societal biases in the data such as the professional gender gap, to biases introduced by the data collection or modeling processes. These biases impact the performance of various components of recommender systems, from offline training, to evaluation and online serving of recommendations in production systems. Specific techniques can help reduce bias at each stage of a recommender system. Reducing bias in our recommender systems is crucial to successfully recommending dream jobs to hundreds of millions members worldwide, while being true to LinkedIn's vision: "To create economic opportunity for *every* member of the global workforce".

## AUDIENCE

This intermediate technical talk is aimed at students, engineers and research scientists interested in understanding and addressing technical challenges that arise in training, evaluating and deploying large-scale recommender systems based on biased real world data.

Students will learn about technical challenges in large-scale real world data and real time applications, beyond idealistic offline datasets often encountered in the academic setting. Industrial practitioners will find practical techniques to address challenges they may have encountered in production systems. Research scientists and academics will find open research problems in recommender systems, and more broadly in fairness and discrimination in AI powered systems. A background in machine learning, data science, data mining or recommender systems is helpful, but not required.

## INTRODUCTION

Machine learning (ML) is at the heart of recommender systems, such as LinkedIn "Jobs You May Be Interested In" (JYMBII), depicted in Figure 1. The goal of JYMBII is to recommend to each member a personalized list of jobs that is relevant to this member. To generate this personalized list of relevant jobs, JYMBII relies on an ML model that has been trained offline to predict the probability that a given member would apply to a given job. This ML model is used online to generate a score measuring the relevance of each job for a given member. These scores are then used to rank jobs by their relevance prior to serving this personalized list online to the member. This process happens in real-time for millions of jobs recommended to LinkedIn's 467+ millions members.

Machine learning models learn what we teach them to learn, which depends on the training data, the model type and complexity, the objective functions in the optimization, and the evaluation metrics. If a machine learning model is trained on biased data, the resulting recommender system may reflect the biases in its recommendations. The goal of this presentation is to share the technical challenges that arise in training, evaluating and deploying a large-scale job recommendation system based on biased real world data. We first describe sources of bias in JYMBII recommender system. We then explain how these different biases can impact the performance of various components of the recommender system. Finally, we propose practical techniques to reduce bias at each stage of a recommender system.

## WHERE BIAS COMES FROM

Biases arise at different stages in a recommender system, from existing societal biases in the data such as the professional gender gap, to biases introduced by the data collection and modeling processes.

**Societal Bias Present in Data:** Societal biases may be inherently present in the data used to train ML models for recommender systems. In a recent study [1], LinkedIn Economists Guy Berger and Alan Fritzler looked at the current state of the professional gender gap. The study reveals that *"there is still a larger gender gap in hiring rates for both leadership and engineering roles"*, as illustrated in Figure 2, courtesy of [1]. A bias in hiring rates also exists across sectors, as shown in Figure 3 reproduced from [1]: In government/education/non-profit and healthcare and pharmaceutical industries, hiring rates are biased towards women, on the contrary to the Technology and Telecommunications sector where hiring rates are biased towards men.

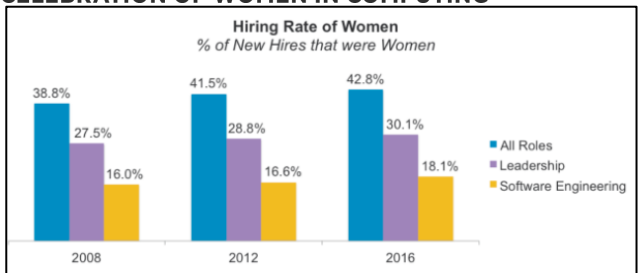

**Figure 2: Women hiring rates evolution (courtesy [1])**

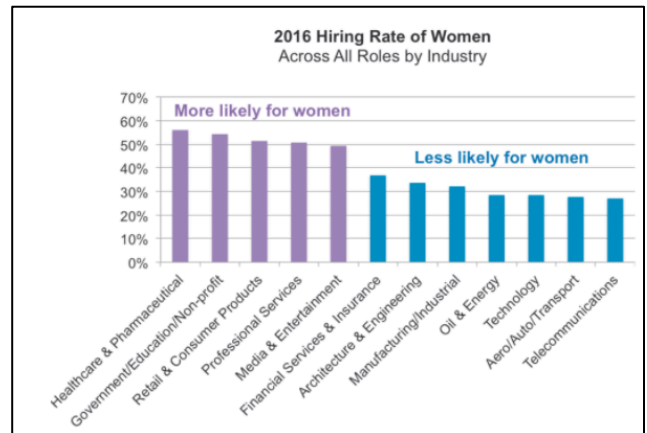

**Figure 3: Women hiring rates across industries (courtesy [1])**

**Bias Introduced by Data Collection:** The data collection process may introduce several types of bias, including:
- *Sample selection bias*, also known as serving bias, arise when data is collected in such a way that some samples are less likely to be included than others. In production systems, data is collected based on a given production model in place. Member feedback such as "click" or "apply" is observed only for items that are displayed, and no feedback is collected on items that are not displayed. Collecting more samples does not reduce this bias.
- *Position bias* is due to members' propensity to click on higher ranked results appearing at the top of the page, rather than results displayed lower on the page.
- *Summary-based presentation bias* happens as members may favor and click on more attractive items, as per their summary (title, url, abstract) or visual components (images, logo).
- *Repetitiveness bias* is due to fatigue from repetitive impressions. Members are less likely to click on item served multiple times in the past, as shown in [2].

**Model Bias:** Modeling bias may arise when the complexity of the model is not appropriately chosen to represent the data. Too simple models with few parameters or insufficient features may suffer from under-fitting [3], and fail to fully represent the outcome they try to model.

## WHY REDUCING BIAS MATTERS

Biases impact the performance of various components of recommender systems, from offline training, to evaluation, to online serving of recommendations in production systems.

**Offline Training:** High bias models usually have unacceptably high training error, and exhibit a small gap between training and test error as shown in Figure 4 from [4]. Increasing the size of the training dataset does not help improve the poor performance of underfitting models.

**Offline Evaluation:** The goal of offline evaluation is to evaluate the performance of a new model on past collected data prior to online launch in production. Only models that demonstrate a good offline performance are selected for online experiments. Offline evaluation allows to iterate on models faster than online experiments, and to minimize the negative impact on members of models that produce irrelevant recommendations. When offline evaluation is conducted on biased data, the offline estimate of the performance may not be accurate, and it may not match the online performance from the experiment. Offline evaluation on biased data may not be a reliable indicator of the online performance of a model.

**Online Recommendations:** A model trained on biased data may reflect implicit biases in its recommendations. For instance, some features may implicitly encode gender and lead an ML model to learn and propagate the professional gender gap in the recommendations it produces. Serving biased recommendations may in turn reinforce societal biases, in a vicious circle.

## HOW TO REDUCE BIAS

Specific techniques can help reduce bias at each stage of a recommender system. We present some techniques that have been used in production recommender and personalized search systems at LinkedIn.

**Online Data Collection:** *Randomized data collection* techniques, such as top K randomization, and fair pairs [5] can help alleviate inherent societal biases in the data as well as sample selection and position biases introduced by the data collection process. *Explore/exploit* techniques such as Multi-Armed Bandits [6,7] are more advanced and economical ways to randomize data collection in order to reduce societal, position and serving biases. They alternate between exploitation phases where items with high predicted score according to the current model are selected, and exploration phases that allow to collect feedback on items with lower predicted score, by randomly selecting items according to a specified distribution. Various exploration schemes exist, such as the epsilon-greedy scheme, the Upper Confidence Bound scheme, or Thomson Sampling.

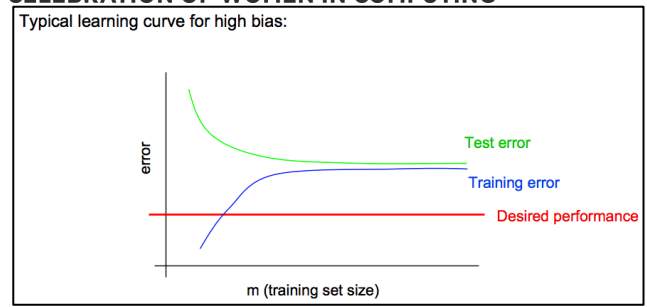

**Figure 4: Learning curve for high bias models (courtesy [4])**

**Offline Modeling and Training:** To reduce bias at the training phase, random negative samples, inferred positives and manually tagged samples can be injected in the training set. Impression discounting can help address repetitiveness bias. Finally sample weighting can be used to compensate for some of the bias. Modeling and correcting for bias in ML models and the results they produce, be it societal biases or biases introduced by data collection or modeling, is an active area of research [9].

**Offline Evaluation:** The replay method [8] is a data driven method that provably produces unbiased offline estimates. It first takes random bucket data as input, and reranks items based on the model to evaluate. It then determines the set of matched impressions, defined as the items who are ranked at the same position by the new ranking model as they were ranked at serve time in the random bucket. The top-K reward estimate (e.g. click-through-rate) is finally computed by counting rewards (e.g. click or apply) on matched impressions.

## CONCLUSION

Recommender systems such as LinkedIn JYMBII rely on ML models to recommend personalized lists of relevant jobs to members. ML models learn what the data teaches them to learn, potentially including biases inherently present in the data or introduced at collection, modeling or serving time. These biases may propagate in each stage of the recommender system and impact their performance and the quality of recommendations. Specific techniques need to be used to effectively reduce bias at each step. Research is still needed in the area of bias modeling and correction in ML, and more broadly on fairness and discrimination in AI powered systems.

## PARTICIPATION STATEMENT

I make a commitment to attend the conference if accepted.

## BIO

Nadia Fawaz is a Staff Software Engineer in Machine Learning/Data Mining at LinkedIn, CA working in the Data Relevance Group. She is technical lead for the job recommendation team, where she oversees deep learning projects. Her research and engineering interests include machine learning for personalization and data privacy. Her work leverages techniques from AI including deep learning, information theory, random matrix theory, statistics and privacy theory, and aims at bridging theory and practice. She has published over 35 papers with over 500 citations, she was a winner of the ACM RecSyS challenge on Context-Aware Movie Recommendations CAMRa2011, and her 2012 UAI paper "Guess Who Rated This Movie: Identifying Users Through Subspace Clustering" was featured in an MIT TechReview article as "The Ultimate Challenge For Recommendation Engines". From 2011 to 2016, she was a principal research scientist at Technicolor research center in Los Altos, CA. From 2009 to 2011, she was a postdoctoral researcher at the Massachusetts Institute of Technology (MIT) Research Laboratory of Electronics (RLE), Cambridge, MA. She received her Ph.D. degree in 2008 and her Diplome d'ingenieur (M.Sc.) in 2005 both in electrical engineering, from Ecole Nationale Superieure des Telecommunications de Paris and EURECOM, France. She is a Member of the IEEE and of the ACM. Web: http://nadiafawaz.com